\documentclass[a4paper]{article}

\usepackage{INTERSPEECH2016}

\usepackage{graphicx}
\usepackage{amssymb,amsmath,bm}
\usepackage{textcomp}
\usepackage{color}
\usepackage{comment}
\usepackage[dvipsnames]{xcolor}

\newcommand{\etal}{\textit{et al.}}

\newcommand{\argmin}{\mathop{\rm arg~min}\limits}
\ninept

\title{Deep Convolutional Neural Networks and Data Augmentation for Acoustic Event Detection}

\makeatletter
\def\name#1{\gdef\@name{#1\\}}
\makeatother \name{{\em Naoya Takahashi$^1$, Michael Gygli$^2$ Beat Pfister$^3$, Luc Van Gool$^4$}}

\address{$^1$Sony Corporation, Japan \\
  $^{2,3,4}$Dept. Information Technology and Electrical Engineering, ETH Zurich, Switzerland \\
  {\small \tt NaoyaA.Takahashi@jp.sony.com, \{gygli, vangool\}@vision.ee.ethz.ch,  pfister@tik.ee.ethz.ch}
}

\begin{document}

  \maketitle
  \begin{abstract}
We propose a novel method for Acoustic Event Detection (AED). In contrast to speech, sounds coming from acoustic events may be produced by a wide variety of sources. Furthermore, distinguishing them often requires analyzing an extended time period due to the lack of a clear sub-word unit. In order to incorporate the long-time frequency structure for AED, we introduce a convolutional neural network (CNN) with a large input field. In contrast to previous works, this enables to train audio event detection end-to-end. 
Our architecture is inspired by the success of VGGNet~\cite{Simonyan2015} and uses small, 3$\times$3 convolutions, but more depth than previous methods in AED.
In order to prevent over-fitting and to take full advantage of the modeling capabilities of our network, we further propose a novel data augmentation method to introduce data variation. Experimental results  show that our CNN significantly outperforms state of the art methods including Bag of Audio Words (BoAW) and classical CNNs, achieving a 16\% absolute improvement.

  \end{abstract}

  \section{Introduction}
    \begin{figure*}[t]
    \centering
    \includegraphics[width=160mm]{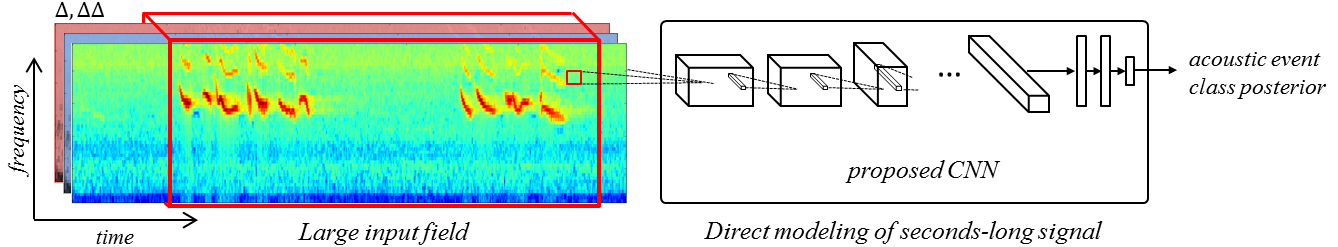}
    \caption{{\it Our deeper CNN models several seconds of acoustic event sound directly and outputs the posterior probability of classes.}}
    \label{fig:overview}
    \end{figure*}

Scenes typically contain many sound sources. While speech is arguably one of the most important types, non-speech sounds such as music or laughter provide important information as well. In most conversations no mention is made of the environment, like its location or people and objects present. Automatic speech recognition (ASR) could benefit from having such contextual knowledge though~\cite{Araki2011}.
Knowing the type of non-speech sounds improves the performance of source separation and speech enhancement~\cite{Ozerov2011}. Furthermore, multi-media tasks such as video classification~\cite{Jiang2011ccv} and video summarization~\cite{li2010multi} have been shown to improve when including audio information. 
Acoustic Event Detection (AED) is attracting more and more attention also due to new applications incl. surveillance~\cite{Radhakrishnan2005,Choi2012,Valenzise2007}, multimedia content retrieval~\cite{Xu2008} and audio segmentation~\cite{Zhuang2010,Zhang2001}. 
    
Traditional methods for AED apply techniques from ASR directly. For instance, Mel Frequency Cepstral Coefficients (MFCC) were modeled with Gaussian Mixture Models (GMM) or Support Vector Machines (SVM) \cite{Eronen2006,Huang2013,Lee2010,Temko2006}. Yet, applying standard ASR approaches leads to inferior performance due to differences between speech and non-speech signals. Thus, more discriminative features were developed. Most were hand-crafted and derived from low-level descriptors such as MFCC~\cite{Phan2015, Pancoast2012}, filter banks~\cite{Choi2015, Beltran2015} or time-frequency descriptors~\cite{Chu2009}. These descriptors are 
frame-by-frame representations (typically frame length is in the order of $ms$) 
and are usually modeled by GMMs to deal with the sounds of entire acoustic events that normally last seconds at least. Another common method to aggregate frame level descriptors is the Bag of Audio Words (BoAW) approach, followed by an SVM~\cite{Pancoast2012, XugangIS2015, HyungjunIS2015,Florian2014}. These models however discard the temporal order of the frame level features, causing considerable information loss. Moreover, methods based on hand-crafted features optimize the feature extraction process and the classification process separately, rather than learning end-to-end.
    
Recently Deep Neural Networks (DNNs) have been very successful at many tasks, including ASR~\cite{Hinton2012,Abdel-hamid2012}, image classification \cite{Alex2012,Simonyan2015}, and visual object detection~\cite{Girshick2014}. One advantage of DNNs is their capability to jointly learn feature representations and appropriate classifiers. Supported by large amounts of training data, more recently, deeper architectures further pushed the state-of-the art for several competitions in computer vision~\cite{Simonyan2015}. In comparison, few AED methods rely on DNNs. One reason is the lack of large, publicly available datasets. In~\cite{Kons2013,Gencoglu2014}, DNNs were built on top of MFCCs. Miquel~\etal~\cite{Espi2015} utilize a Convolutional Neural Network (CNN)~\cite{LeCun1998} to extract features from spectrograms. These networks are still relatively shallow (e.g. 3 layers). Furthermore, the networks take only a few frames as input and the complete acoustic events are modeled by Hidden Markov Models (HMM) or simply by calculating the mean of the network outputs, which is too simple to model complicated acoustic event structures.
    
In this work, we introduce novel network architectures with up to 9 layers and a large input field. The large input field allows the networks to directly model entire audio events and be trained end-to-end, as depicted in {\it Fig. \ref{fig:overview}}. 
Our network architecture is inspired by “VGG Net”~\cite{Simonyan2015} which obtained second place in the ImageNet 2014 competition and was successfully applied for ASRs \cite{Sercu2015}.
The main idea of VGG Net is to replace large (typically 9$\times$9) convolutional kernels by a stack of 3$\times$3 kernels without pooling between these layers. Advantages of this architecture are (1) additional non-linearity hence more expressive power,
and (2) a reduced number of parameters (i.e. one 9$\times$9 convolution layer with $C$ channel has $9^2C^2=81C^2$ weights while three-layer 3$\times$3 convolution stack has $3(3^2C^2)=27C^2$ weights). 
Our first goal is to adapt the VGG Net architecture to AED. In order to train our network we further propose a novel data augmentation method, especially effective for AED.
For our experiments, we created a new dataset harvested from the Freesound repository~\cite{freesound}
and conducted acoustic event classification. Experimental results show that our deeper CNN significantly outperforms several baseline techniques, including state-of-the-art methods based on BoAW and classical DNNs. We further show that the proposed data augmentation method improves the performance by more than 12\%.

\section{Architectural and Training Novelties}

\subsection{Convolutional Network Architecture}
    
We propose two CNN architectures, adapted to AED, as outlined in {\it Table~\ref{tab:cnnarch}}. 
Architecture $A$ has 4 convolutional and 3 fully connected layers, while Architecture $B$ has 9 weight layers: 6 convolutional and 3 fully connected. 
In this table, the convolutional layers are described as conv({input feature maps}, {output feature maps}). All convolutional layers have 3$\times$3 kernels, thus henceforth kernel size is omitted. The convolution stride is fixed to 1. The max-pooling layers are indicated as $time \times frequency$ in Table~\ref{tab:cnnarch}. They have a stride equal to the pool size.
All hidden layers except the last fully-connected layer are equipped with the Rectified Linear Unit (ReLU) non-linearity.
In contrast to \cite{Simonyan2015}, we do not apply zero padding before convolution since the output size of the last pooling layer is still large enough in our case.
The networks were trained by minimizing the cross entropy loss $L$ with $l_1$ regularization using back propagation:
      \begin{equation}
        \argmin_{W} \sum_i L(X_i,Y_i,W)+\lambda\|W\|_1
        \label{eq:obj}
      \end{equation}
where $X_i,Y_i,W$ are the $i$th input, label and network parameters, respectively. $\lambda$ is a constant which is set to $10^{-6}$ in this work.
      \setlength{\tabcolsep}{1.2mm} 
      \begin{table}[t]
        \caption{\label{tab:cnnarch} {\it The architecture of our deeper CNNs. Unless mentioned explicitly convolution layers have 3$\times$3 kernels.}}
        \vspace{2mm}
        \centerline{
          \footnotesize
          \begin{tabular}{ c | c | c | c | c } 
            \hline
            \multicolumn{1}{c|}{} & \multicolumn{2}{c|}{Baseline} & \multicolumn{2}{c}{Proposed CNN}\\
            \hline
            \#Fmap & DNN & Classic CNN &	A          & B \\
            \hline
            64   &      & conv5$\times$5 (3,64) & conv(3,64)   & conv(3,64)  \\
            	 &	    & pool 1$\times$3	     & conv(64,64)  & conv(64,64) \\
                 &      & conv5$\times$5(64,64)& pool 1$\times$2	    & pool 1$\times$2    \\
            \hline                 
            128  &      &                & conv(64,128) & conv(64,128) \\
                 &      &                & conv(128,128)& conv(128,128) \\
                 &      &                & pool 2$\times$2	    & pool 2$\times$2 \\
            \hline
            256  &      &                &              & conv(128,256) \\
                 &      &                &              & conv(128,256) \\
                 &      &                &              & pool 2$\times$1 \\
            \hline
            FC   & FC4096 &        &        &  \\
                 & FC2048 &  FC1024      &  FC1024      & FC2048 \\
                 & FC2048 &  FC1024      &  FC1024      & FC2048 \\
                 & FC28   &  FC28        &  FC28        & FC28 \\
            \hline
            \multicolumn{1}{c|}{} & \multicolumn{4}{c}{softmax} \\
            \hline
            \hline
            \#param & 258$\times 10^6$  & 284$\times 10^6$ & 233$\times 10^6$   & 257$\times 10^6$ \\
            \hline
          \end{tabular}
        }
      \end{table}

\subsection{Large input field}

In ASR, few-frames descriptors are typically concatenated and modeled by GMM or DNN~\cite{Hinton2012,Abdel-hamid2012}. This is reasonable since they aim to model sub-word units like phonemes which typically last less than a few hundreds of $ms$. The sequence of sub-word units is typically modeled by a HMM. Most works in AED also follow similar frameworks, where signals lasting from tens to hundreds of $ms$ are modeled first. These small input field representations are then aggregated to model longer signals by HMM, GMM~\cite{Zhuang2010,Eronen2006,Espi2015,Xu2008,Deng2014} or a combination of BoAW and SVM~\cite{XugangIS2015, HyungjunIS2015,Florian2014}. Yet, unlike speech signals, non-speech signals are much more diverse even within a category and it is not clear that a sub-word approach is suitable for AED. Hence, we design a network architecture that directly models the entire acoustic event, based on a single input of multiple seconds. This also enables the networks to optimize the parameter end-to-end.

\subsection{Data Augmentation}

Since the proposed CNN architectures 
have many hidden layers and a large input field, the number of parameters is high, as shown in the last row of {\it Table~\ref{tab:cnnarch}}. A large number of training data is vital to train such networks. Jaitly~\etal~\cite{Jaitly2013} showed that the data augmentation based on Vocal Tract Length Perturbation (VTLP) is effective to improve ASR performance. VTLP attempts to alter the vocal tract length during extraction of descriptors, such as a log filter bank, and perturbs the data in a certain non-linear way.
\\   
In order to introduce more data variation, we propose a different augmentation technique.
For most sounds coming with an acoustic event, mixed sounds from the same class also belong to that class, except when the class is differentiated by the number of sound sources. For example, when mixing two different ocean surf sounds, or of breaking glass, or birds tweeting, the result still belongs to the same class. Considering this property we produce augmented sounds by randomly mixing two sounds of a class, with randomly selected timings. In addition to mixing sounds, we further perturb the sound by moderately modifying frequency characteristics of each source sound by boosting/attenuating a particular frequency band. An augmented data sample $s_{aug}$ is generated from source signals for the same class as the one both $s_1$ and $s_2$ belong to, as follows:
      \begin{equation}
        s_{aug} = \alpha \Phi(s_1(t),\psi_1) + (1-\alpha)\Phi(s_2(t-\beta T), \psi_2)
        \label{eq1}
      \end{equation}
where $ \alpha, \beta \in [0,1)$ are uniformly distributed random values, $T$ is the maximum delay and $\Phi(\cdot,\psi)$ is an equalizing function parametrized by $\psi$. In this work, we used a second order parametric equalizer parametrized by $\psi=(f_0,g,Q)$ where $f_0 \in [100,6000]$ is the center frequency, $g \in [-8,8]$ is a gain and $Q \in [1,9]$ is a Q-factor. An arbitrary number of such synthetic samples can be obtained by randomly selecting parameters $\alpha, \beta, \psi$ for each data augmentation.
We refer to this approach as Equalized Mixture Data Augmentation (EMDA).
      
      \begin{figure}[t]
        \centering
        \includegraphics[width=75mm]{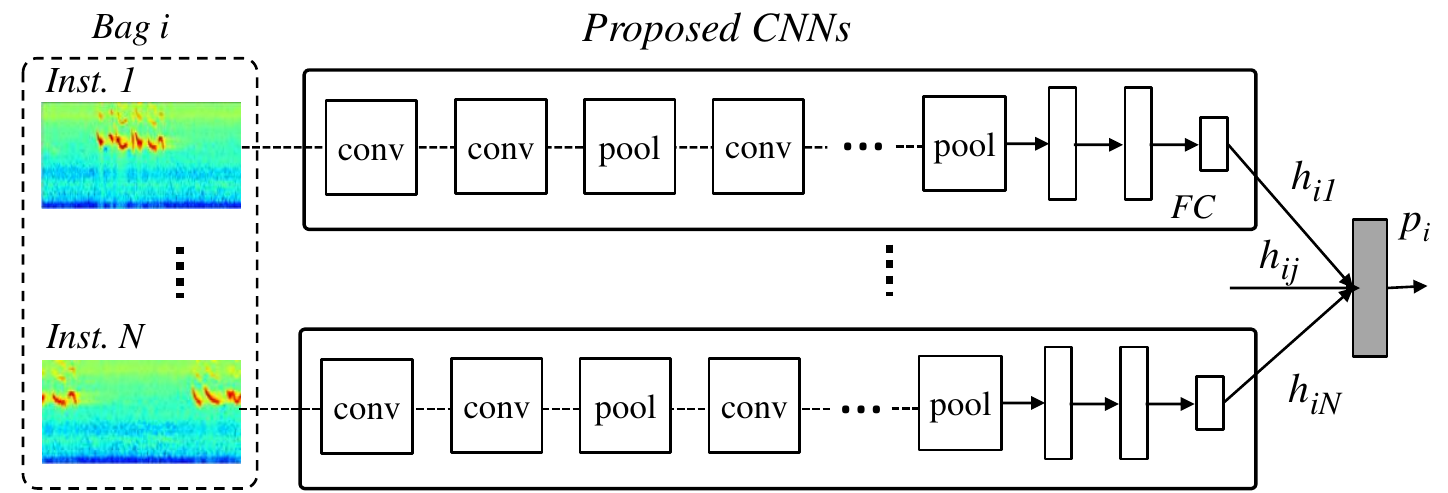}
        \caption{{\it Architecture of our deeper CNN model adapted to MIL. The softmax layer is replaced with the aggregation layer}.} 
        \label{fig:mil}
      \end{figure}
    
\subsection{Multiple Instance Learning}

Since we used web data to build our dataset (see {\it Sec. \ref{sec:database}}), the training data is expected to be noisy and to contain outliers.
In order to alleviate the negative effects of outliers, we also employed multiple instance learning (MIL) \cite{Wu2015,Zhang2005}. In MIL, data is organized as bags $\{X_i\}$ and within each bag there are a number of instances $\{x_{ij}\}$. Labels $\{Y_i\}$ are provided only at the bag level, while labels of instances $\{y_{ij}\}$ are unknown. A positive bag means that at least one instance in the bag is positive, while a negative bag means that all instances in the bag are negative. We adapted our CNN architecture for MIL as shown in {\it Fig.~\ref{fig:mil}}. $N$ instances $\{x_1,\cdots,x_N\}$ in a bag are fed to a replicated CNN which shares parameters. The last softmax layer is replaced with an aggregation layer where the outputs from each network $h = \{h_{ij}\} \in R^{M\times N}$ are aggregated. Here, $M$ is the number of classes. The distribution of class of bag $p_i$ is calculated as
$p_i = f(h_{i1},h_{i2},\cdots,h_{iN})$
where $f()$ is an aggregation function. In this work, we investigate 2 aggregation functions: max aggregation
      \begin{eqnarray}
        \label{eq:max}
        p_i = \cfrac{exp(\hat{h_i})}{\sum_i exp(\hat{h_i})}\\
        \hat{h_i} = \max_{j}(h_{ij})
      \end{eqnarray}
      and Noisy OR aggregation \cite{David1989},
      \begin{eqnarray}
        p_i = 1- \prod_j (1-p_{ij})\\
        p_{ij} = \cfrac{exp(h_{ij})}{\sum_j exp(h_{ij})}.
        \label{eq:nor}
      \end{eqnarray}
Since it is unknown which sample is an outlier, we can not be sure that a bag has at least one positive instance. However, the probability that all instances in a bag are negative exponentially decreases with $N$, thus the assumption becomes very realistic.

\section{Experiments}
    
\subsection{Dataset}
\label{sec:database}

The proposed methods are evaluated on a novel acoustic event classification database \footnote{The dataset is available at https://data.vision.ee.ethz.ch/cvl/ae\_dataset} harvested from Freesound \cite{freesound}, which is a repository of audio samples uploaded by users. The database consists of 28 events as described in {\it Table \ref{tab:database}}. Note that since the sounds in the repository are tagged in free-form style and the words used vary a lot, the harvested sounds contain irrelevant sounds. For instance, a sound tagged 'cat' sometime does not contain a real cat meow, but instead a musical sound produced by a synthesizer. Furthermore sounds were recorded with various devices under various conditions (e.g. some sounds are very noisy and in others the acoustic event occurs during a short time interval among longer silences). This makes our database more challenging than previous datasets such as \cite{Nakamura2000}.

In order to reduce the noisiness of the data, we first normalized the harvested sounds and eliminated silent parts.
If a sound was longer than 12 sec, we split the sound in pieces so that split sounds were less than 12 sec. 
All audio samples were converted to 16 kHz sampling rate, 16 bits/sample, mono channel. Similar to \cite{Deng2014}, the data was randomly split into training set (75\%) and test set (25\%). Only the test set was manually checked and irrelevant sounds not containing the target acoustic event, were omitted. The data augmentation was applied only to the training set.
        \setlength{\tabcolsep}{1.6mm} 
        \begin{table}[t]
            \caption{\label{tab:database} {\it The statistics of the dataset.}}
            \vspace{2mm}
            \centering{
              \footnotesize
              \begin{tabular}{ c | p{9mm} | c | c | p{9mm} | c } 
                \hline
                Class &	Total minutes	& \# clip & Class	& Total minutes	& \# clip \\
                \hline
                Acoustic guitar	&	23.4	&	190	&	Hammer	&	42.5	&	240	\\
                Airplane	    &	37.9	&	198	&	Helicopter	&	22.1	&	111	\\
                Applause	    &	41.6	&	278	&	Knock	&	10.4	&	108	\\
                Bird        	&	46.3	&	265	&	Laughter	&	24.7	&	201	\\
                Car	            &	38.5	&	231	&	Mouse click	&	14.6	&	96	\\
                Cat	            &	21.3	&	164	&	Ocean surf	&	42	&	218	\\
                Child	        &	19.5	&	115	&	Rustle	&	22.8	&	184	\\
                Church bell	    &	11.8	&	71	&	Scream	&	5.3	&	59	\\
                Crowd	        &	64.6	&	328	&	Speech	&	18.3	&	279	\\
                Dog barking	    &	9.2	    &	113	&	Squeak	&	19.8	&	173	\\
                Engine	        &	47.8	&	263	&	Tone	&	14.1	&	155	\\
                Fireworks	    &	43	    &	271	&	Violin	&	16.1	&	162	\\
                Footstep	    &	70.3	&	378	&	Water tap	&	30.2	&	208	\\
                Glass breaking	&	4.3	    &	86	&	Whistle	&	6	&	78	\\
                \hline
                \multicolumn{3}{c|}{} & \multicolumn{1}{c|}{\textbf{Total}} & 
                    \multicolumn{1}{c|}{768.4} & \multicolumn{1}{c}{5223}  \\
                \hline
              \end{tabular}
            }
        \end{table}

\subsection{Implementation details}

Through all experiments, 49 band log-filter banks, log-energy and their delta and delta-delta were used as a low-level descriptor using 25 ms frames with 10 ms shift, except for the BoAW baseline described in {\it Sec.~\ref{sec:ex1}}. Input patch length was set to 400 frames (i.e. 4 sec). The effects of this length were further investigated in {\it Sec.~\ref{sec:ex2}}. During training, we randomly crop 4 sec for each sample.
The networks were trained using mini-batch gradient descent based on back propagation with momentum.
We applied dropout \cite{Hinton2012do} to each fully-connected layer with keeping probability $0.5$. The batch size was set to 128, the momentum to 0.9. For data augmentation we used VTLP and the proposed EMDA. The number of augmented samples is balanced for each class. 
During testing, 4 sec patches with 50\% shift were extracted and used as input to the Neural Networks. The class with the highest probability was considered the detected class. The models were implemented using the Lasagne library \cite{lasagne}.

\subsection{Experimental Results and Discussions}

\subsubsection{State-of-the-art comparison}
\label{sec:ex1}
In our first set of experiments we compared our proposed deeper CNN architectures  to three different state-of-the-art baselines, namely, BoAW~\cite{Pancoast2012}, 
HMM+DNN/CNN as in~\cite{Gencoglu2014}, and classical DNN/CNN with large input field.
\vspace{2mm}
\\
\textbf{BoAW} \hspace{2mm} We used MFCC with delta and delta-delta as a low-level descriptor.
K-means clustering was applied to generate an audio word code book with 1000 centers. We evaluated both SVM with a $ \chi^2$ kernel and a 4 layer DNN as a classifier. The layer sizes of the DNN classifier were (1024, 256, 128, 28).
\\
\textbf{DNN/CNN+HMM} \hspace{2mm} We evaluate the DNN-HMM system. The neural network architectures are described in the left 2 columns in Table \ref{tab:cnnarch}. Both DNN and CNN models are trained to estimate HMM state posteriors. The HMM topology consists of one state per acoustic event, and an ergodic architecture in which all states have a self-transition and equal transitions to all other states, as in \cite{Espi2015}. The input patch length for CNN, DNN is 30 frames with 50\% shift.  
\\
\textbf{DNN/CNN+Large input field} \hspace{2mm} In order to evaluate the effect of using the proposed CNN architectures, we also evaluated the baseline DNN/CNN architectures with the same large input field, namely, 400 frame patches. 
\\
The classification accuracies of these systems trained with and without data augmentation are shown in Table \ref{tab:ex1}. Even without data augmentation, the proposed CNN architectures outperform all previous methods. Furthermore, the performance is significantly improved by applying data augmentation, achieving 12.5\% improvement for the $B$ architecture. The best result was obtained by the $B$ architecture with data augmentation. 
It is important to note that the $B$ architecture outperforms classical DNN/CNN even though it has less parameters as shown in {\it Table \ref{tab:cnnarch}}. This result supports the efficiency of deeper CNNs with small kernels for modelling large input fields.

        \begin{table}[t]
            \caption{\label{tab:ex1} {\it Accuracy of the deeper CNN and baseline methods, trained with and without data augmentation (\%).}}
            \vspace{2mm}
            \centering{
              \begin{tabular}{ c | c  c } 
                \hline
                \multicolumn{1}{c|}{} & \multicolumn{2}{c}{Data augmentation}\\
                Method      &	without	&	with	\\
                \hline\hline
                BoAW+SVM	&	74.7	&	79.6	\\
                BoAW+DNN	&	76.1	&	80.6	\\
                \hline
                DNN+HMM	    &	54.6	&	75.6	\\
                CNN+HMM     &	67.4	&	86.1	\\
                \hline
                DNN+Large input	& 62.0	&	77.8	\\
                CNN+Large input & 77.6	&	90.9	\\
                \hline
                $A$	        &	77.9	&	91.7	\\
                $B$	        &	80.3	    & \textbf{92.8}	\\
                \hline
              \end{tabular}
            }
        \end{table}        

\subsubsection{Effectiveness of large input field}
\label{sec:ex2}

\begin{figure}[t]
        \centering
        \includegraphics[width=77mm]{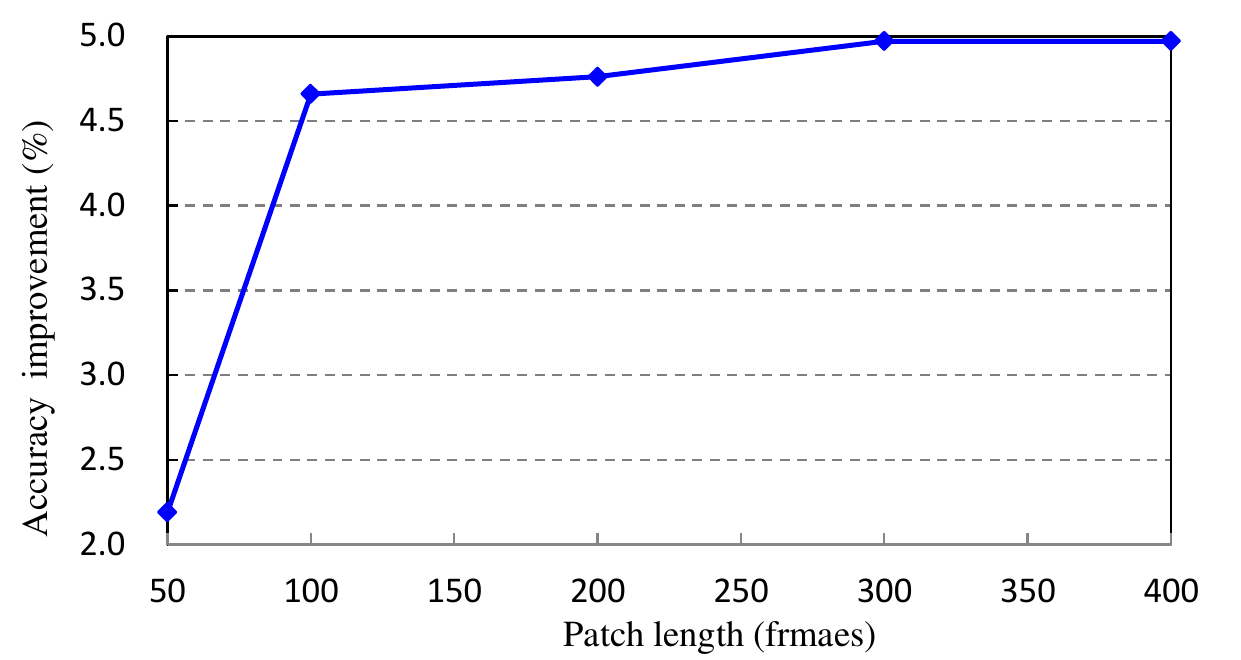}
        \caption{{\it Performance of our network for different input patch lengths. The plot shows the increase over using a CNN+HMM with a small input field of $30$ frames.}}
        \label{fig:ex2}
      \end{figure}

Our second set of experiments focuses on input field size. We tested our CNN with different patch size {50, 100, 200, 300, 400} frames (i.e. from 0.5 to 4 sec). The $B$ architecture was used for this experiment. As a baseline we evaluated the CNN+HNN system described in {\it Sec.~\ref{sec:ex1}} but using our architecture $B$, rather than a classical CNN.  
The performance improvement over the baseline is shown in {\it Fig.~\ref{fig:ex2}}. The result shows that larger input fields improve the performance. Especially the performance with patch length less than 1 sec sharply drops. This proves that modeling long signals directly with deeper CNN is superior to handling long sequences with HMMs.
        
\subsubsection{Effectiveness of data augmentation}

\begin{figure}[t]
        \centering
        \includegraphics[width=\linewidth]{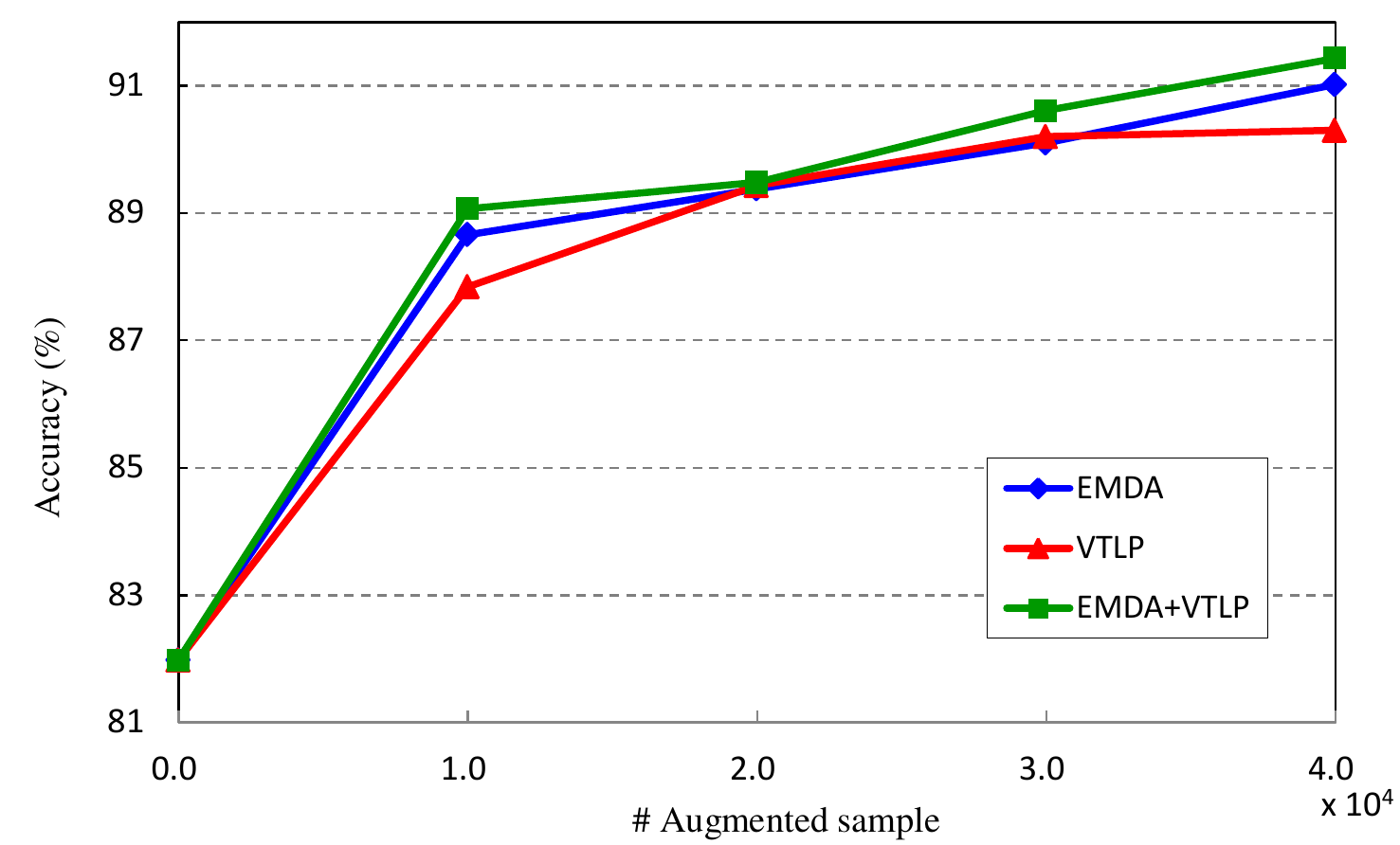}
        \caption{{\it Effects of different data augmentation methods with varying amounts of augmented data.}}
        \label{fig:ex3}
      \end{figure}

We verified the effectiveness of our EMDA data augmentation method in more detail. We evaluated 3 types of data augmentation: EMDA only, VTLP only, and a mixture of EMDA and VTLP (50\%, 50\%) with different numbers of augmented sample {10k, 20k, 30k, 40k}. Fig.~\ref{fig:ex3} shows that using both EDMA and VTLP always outperforms EDMA or VTLP only. 
This shows that EDMA and VTLP perturbs original data and create new samples in a different way, providing more effective variation of data and helping to train the network to learn a more robust and general model from limited amount of data. 
        
\subsubsection{Effects of Multiple Instance Learning}

Finally, the $A$ and $B$ architectures with a large input field were adapted to MIL to handle the noise in the database. The number of parameters were identical since both max and Noisy OR aggregation methods are parameter free. The number of instances in a bag was set to 2. We randomly picked 2 instances from the same class during each epoch of the training.  
Table \ref{tab:mil} shows that MIL didn't improve performance.
However, MIL with a medium size input field (i.e. 2 sec) performs as good as or even slightly better than single instance learning with a large input field. This is perhaps due to the fact that the MIL took the same size input length (2 sec $\times$2 instances $ = $ 4 sec), while it had less parameter. Thus it managed to learn a more robust model.
      
        \begin{table}[h]
            \footnotesize
            \caption{\label{tab:mil} {\it Accuracy of MIL and normal training (\%).}}
            \vspace{1mm}
            \centering{
              \begin{tabular}{ c | c | c c c}
                \hline
                            &  Single         & \multicolumn{3}{c}{MIL}\\
                Architecture & instance	& Noisy OR  & Max & Max (2sec)	\\
                \hline
                $A$	        &	91.7    & 90.4      & 92.6  & \textbf{92.9}	\\
                $B$	        &	92.8    & 91.3      & 92.4  & 92.8  \\
                \hline
              \end{tabular}
            }
        \end{table}  
        
\section{Conclusions}

We proposed new CNN architectures and showed that they allow to learn a model for AED end-to-end, by directly modeling a several seconds long signal.
We further proposed a method for data augmentation that prevents over-fitting and leads to superior performance even when training data is fairly limited.
Experimental results shows that proposed methods significantly outperforms state of the arts.
We further validated the effectiveness of deeper architectures, large input fields and data augmentation one by one.  
Future work will be directed towards applying the proposed AED to different applications such as video segmentation and summarization.

  \newpage
  \eightpt
  \bibliographystyle{IEEEtran}

  \bibliography{mybib}
\end{document}